\documentclass{ws-mplb}

\begin{document}

\markboth{Sergey L. Bud'ko}{Superconductivity in K- and Na- doped BaFe$_2$As$_2$...}

%
\catchline{}{}{}{}{}
%

\title{SUPERCONDUCTIVITY IN K- AND Na- DOPED BaFe$_2$As$_2$: WHAT CAN WE LEARN FROM HEAT CAPACITY AND PRESSURE DEPENDENCE OF T$_c$}

\author{SERGEY L. BUD'KO}

\address{Ames Laboratory, US DOE and Department of Physics and Astronomy, Iowa State University\\
Ames, Iowa 50011, USA\\
budko@iastate.edu}

\maketitle

\begin{history}
\received{(Day Month Year)}
\revised{(Day Month Year)}
\end{history}

\begin{abstract}

A brief overview of changes in the superconducting transition temperature under pressure and evolution of specific heat capacity jump at $T_c$ for two related families of iron - based superconductors,  Ba$_{1-x}$K$_x$Fe$_2$As$_2$ ($0.2 \leq x \leq 1.0$) and   Ba$_{1-x}$Na$_x$Fe$_2$As$_2$ ($0.2 \leq x \leq≤ 0.9$) will be given. For  Ba$_{1-x}$K$_x$Fe$_2$As$_2$ the specific heat capacity jump at $T_c$ measured over the whole extent of the superconducting dome shows clear deviation from the empirical, $\Delta C_p (T_c) \propto T_c^3$, scaling (known as the BNC scaling) for $x > 0.7$.  At the same concentrations range apparent equivalence of effects of  pressure and K- substitution on $T_c$ fails. These observations suggests a significant change of the superconducting state for $x > 0.7$. In contrast, the data for the large portion of  Ba$_{1-x}$Na$_x$Fe$_2$As$_2$ ($0.2 \leq x \leq 0.9$) series follow the BNC scaling. However, the pressure dependence of $T_c$ (measured up to $\sim 12$ kbar) have clear non-linearities for Na concentration in 0.2 - 0.25 region, that may be consistent with $T_c$ crossing the phase boundaries of the emergent, narrow, antiferromagnetic/tetragonal phase. Results will be discussed in context other studies of these two and related families of iron-based superconductors.

\end{abstract}

\keywords{iron pnictide superconductors; pressure; heat capacity.}

\section{Introduction}

Of many families of Fe-based superconductors and related materials \cite{joh10a,ste11a,joh11a,joh11b,wan12a}  the {\it AE}Fe$_2$As$_2$ ({\it AE} = alkaline earth and Eu), so called 122 family, is apparently the most studied \cite{can10a,man10a,nin11a}. The 122 family allows for substitution on all three crystallographic sites and, as a result, a complex combination of carrier-doping and anisotropic steric effects can be studied, while maintaining, at room temperature and ambient pressure, the same, tetragonal, ThCr$_2$Si$_2$ - type crystal structure. Starting from the parent,  {\it AE}Fe$_2$As$_2$ compounds, superconductivity can be induced either by properly designed substitution, or by application of pressure, or by combination of both routes. 

Historically, the main body of work on the 122 family was focused on the transition metal ({\it TM}) substitutions for Fe: {it AE}(Fe$_{1-x}${\it TM}$_x$)$_2$As$_2$, due to relative ease of growing homogeneous, relative large single crystals \cite{can10a,nin11a}. Substitutions for $AE$, as in the Ba$_{1-x}$K$_x$Fe$_2$As$_2$ series \cite{rot08a,rot08b,che09a,avc12a},  or for As as in the BaFe$_2$(As$_{1-x}$P$_x$)$_2$ series \cite{jia09a,kas10a} have been explored, but both series require significant efforts to achieve homogeneity and/or reasonable size of the crystals. 

The global phase diagrams obtained by substitution for Ba [e.g. hole-doping in Ba$_{1-x}$K$_x$Fe$_2$As$_2$], Fe [e.g. electron-doping in Ba(Fe$_{1-x}$Co$_x$)$_2$As$_2$] or As [e.g.isovalent substitution in BaFe$_2$(As$_{1-x}$P$_x$)$_2$] in BaFe$_2$As$_2$, as well as application of pressure \cite{col09a}, to this parent compound appear to be very similar. \cite{pag10a} First, the temperature of structural and magnetic transitions decreases, then superconductivity emerges with a region of coexistence of superconductivity and antiferromagnetism. On further substitution (or under higher pressure) the  magnetic and structural transitions are suppressed, the superconducting transition temperature passes through the maximum  and gradually goes to zero, or to a small finite value as in the case of complete substitution of K for Ba, in KFe$_2$As$_2$. Closer examination of the globally similar phase diagrams though, point to clear differences in details that allow to gain an insight into the complex physics of these materials \cite{ste11a,can10a,man10a,nin11a}.

The results of two sets of measurements on a large, specific selection of samples will be discsussed below. The first set is comprised of the initial ($P \lesssim 12$ kbar) pressure dependence of the superconducting transition temperature, $T_c(P)$. The second set consists of of the data on the evolution of the jump in temperature-dependent specific heat capacity at the superconducting transition, $\Delta C_p (T_c)$. 

Measurement of  $T_c$ under pressure is one of the traditional experiments performed on superconductors \cite{bra6569a,lor05a,sch07a,chu09a,sch13a}. At a minimum, such data allow for evaluation of possible equivalence of pressure and chemical substitution that was suggested for several 122 series \cite{kim09a,dro10a,kli10a,kim11a}.  Moreover, under favorable circumstances such a dataset can shed light or make conjectures on the details of the electronic structure and mechanism of superconductivity, like Lifshitz transitions \cite{lif60a}, pressure induced oxygen ordering effects, change of the symmetry of the superconducting pairing, or evolution of the superconducting gap structure \cite{mak65a,xio92a,cao95a,fie96a,taf13a,taf14a,tau14a}. 

The experimental data on the jump in specific heat capacity at the superconducting transition are usually analyzed  in terms of $\Delta C_p (T_c)/\gamma T_c$, where $\gamma$ is a normal state Sommerfeld coefficient of the material. In the weak coupling BCS limit $\Delta C_p (T_c)/\gamma T_c = 1.43$. This value can go as high as $\approx 3$ in the strong coupling limit \cite{car90a}. A lot of work was done on analysis of $\Delta C_p (T_c)/\gamma T_c$ in the cases of anisotropic superconducting order parameter, multiband superconductivity, and superconducting materials with magnetic or with Kondo impurities \cite{bou01a,joh13a,ska64a,mul72a,map76a,shi73a,ope04a}. Such analysis becomes difficult if not impossible in the case of iron-based superconductors. Whereas $\Delta C_p (T_c)$ and $T_c$ are easy to measure (in homogeneous samples), to evaluate normal state $\gamma$ in superconductors, usually either specific heat capacity measurements are performed down to low temperatures in magnetic field in excess of (upper) critical field (assuming that normal state $\gamma$ does not depend on magnetic field), or different contributions to a zero field specific heat capacity are carefully analized at temperatures above $T_c$ and the electronic normal state contribution is separated.  Since superconducting transition temperatures and upper critical fields in the iron-based superconductors are high \cite{joh10a,ste11a,can10a,gur11a}, and ambiguous magnetic contribution to specific heat capacity is often present, a reliable evaluation of the normal state $\gamma$ in these materials is difficult to achieve.

In lieu of reliable data on $\gamma$ in Fe-based superconductors, an alternatime approach to analize specific heat capacity data emerged. It was observed that many Fe-based 122 superonductors follow the empirical trend, suggested in Ref. \refcite{bud09b} and expanded in Refs. \refcite{kim11b,kim12a}, the so-called BNC scaling, $\Delta C_p (T_c) \propto T_c^3$.  It was proposed (even if on a limited dataset) that for elemental superconductors, and representative A-15 and heavy fermion compounds  $\Delta C_p (T_c) \propto T_c^n$ scaling with different, $n \approx 1.7 - 2$, takes place  \cite{kim11a}. Several theoretical models were used to explain the BNC scaling in iron - pnictides. The $\Delta C_p (T_c) \propto T_c^3$ behavior is expected for a quantum critical metal undergoing a pairing instability \cite{zaa09a}; in the limit of strong pair-breaking within BCS theory for $d$- or $s_{\pm}$- symmetry of the superconducting gap \cite{kog09a}; or as a result of an interplay between superconductivity and magnetism \cite{vav11a}. Rather than discussing {\it pro et contra} arguments for different theories, here we will treat the BNC scaling as an experimental observation valid for many Fe-As superconductor, with no association with one or another specific theoretical paradigm. 

\begin{figure} [ht]
\vspace*{16pt}
\centerline{\psfig{file=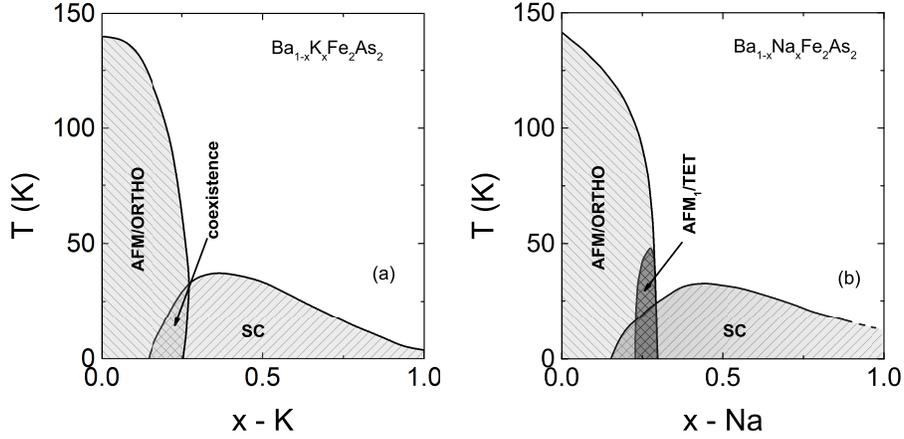,width=12cm}}
\vspace*{8pt}
\caption{Schematic {\it x - T} phase diagrams for the  Ba$_{1-x}$K$_x$Fe$_2$As$_2$ and  Ba$_{1-x}$Na$_x$Fe$_2$As$_2$ series (after Refs. ~\protect\refcite{rot08b,avc12a,avc13a,avc14a}). Labels: AFM/ORTHO - antiferromagnetic / orthorhombic phase, AFM$_1$/TET - antiferromagnetic / tetragonal phase , SC - superconducting region. \label{f1}}
\end{figure}

Two families that will be briefly discussed here are Ba$_{1-x}$K$_x$Fe$_2$As$_2$ and  Ba$_{1-x}$Na$_x$Fe$_2$As$_2$. In both cases the alkali metal substitution in place of Ba provides hole-doping to the system.  It was described \cite{cor10a} that the average structural changes in the FeAs layers of  Ba$_{1-x}${\it A}$_x$Fe$_2$As$_2$ ({\it A} = alkali metal)  are independent of the average radius of the alkal metal cation {\it A} and the mismatch in radii between the alkaline earth and alkali metal cations, and result from decreasing the electron count. A quantitative similarity between the evolution of the competing antiferromagnetic order and superconductivity in the Ba$_{1-x}$K$_x$Fe$_2$As$_2$ and  Ba$_{1-x}$Na$_x$Fe$_2$As$_2$ series was reported. Indeed, the {\it x - T} phase diagrams for these families (based on \cite{rot08b,avc12a,avc13a,avc14a}) schematically presented in Fig. \ref{f1}, bear great resemblance.

The  Ba$_{1-x}$K$_x$Fe$_2$As$_2$ series is rather well studied. In these series superconductivity is observed over a wide range of K - concentrations, $0.15 \lesssim x \leq 1.0$ \cite{rot08b,che09a,avc12a}  (as compared to $0.03 \lesssim x \lesssim 0.15$ for the electron-doped Ba(Fe$_{1-x}$Co$_x$)$_2$As$_2$ \cite{nin08a}).  For  underdoped Ba$_{1-x}$K$_x$Fe$_2$As$_2$ superconductivity and magnetism microscopically co-exist \cite{wie11a} [similarly to what was observed in the electron-doped Ba(Fe$_{1-x}${\it TM}$_x$)$_2$As$_2$ \cite{can10a,pra09a}]. Near optimal doping, $x \sim 0.4$, several experiments and theoretical calculations give evidence for a nodeless, near constant, $s_\pm$ superconducting gap that changes sign between hole and electron pockets. \cite{din09a,chr08a,luo09a,gra09a,maz09a} Recent ARPES measurements \cite{nak11a} suggest that nearly isotropic  $s_\pm$ superconducting gap exists in a wide doping range, $0.25 \leq x \leq 0.7$. It was proposed \cite{ota14a} that for significantly higher K-doping, $0.076 \leq x \leq 0.93$  the superconducting gap is still of  $s_\pm$ symmetry but with a significant anisotropy.

KFe$_2$As$_2$ stands out among the members of the Ba$_{1-x}$K$_x$Fe$_2$As$_2$ series. The reported Fermi surface of KFe$_2$As$_2$ differs from that of the optimally doped  Ba$_{1-x}$K$_x$Fe$_2$As$_2$ having three hole pockets, two centered at the $\Gamma$ point in the Brillouin zone, and one around the $M$ point \cite{sat09a} with no electron pockets. Quantum criticality and nodal or $d$-wave superconductivity in KFe$_2$As$_2$ was suggested in a number of publications. \cite{has10a,don10a,ter10a,don10b,rei12a,mai12a,tho11a,oka12a}  Possible evolution from $s_\pm$ to $d$-wave in the  Ba$_{1-x}$K$_x$Fe$_2$As$_2$ series has also been discussed \cite{rei12b,hir12a}.

Na - substitution for the $AE$ in 122 family  appears to be  less explored. Reports of superconductivity induced by Na-substitution in CaFe$_2$As$_2$ appeared fairly early  \cite{shi08a,wug08a}, with a tentative $x - T$ phase diagram for the Ca$_{1-x}$Na$_x$Fe$_2$As$_2$ series  published a few years later \cite{zha11a}. A Sr$_{1-x}$Na$_x$Fe$_2$As$_2$ sample with $T_c \approx 35$ K was studied in Ref. \refcite{gok09a}, and later the physical properties in the Sr$_{1-x}$Na$_x$Fe$_2$As$_2$ series for a $x \leq 0.4$ were presented. \cite{cor11a}

The Ba$_{1-x}$Na$_x$Fe$_2$As$_2$ series, where Na is substituted for Ba, offers an almost complete range of substitution \cite{cor10a,avc13a}. One of the complications for this series is that its end member, NaFe$_2$As$_2$ ($T_c \sim 11 - 12$ K), was reported to be metastable.  NaFe$_2$As$_2$ cannot be formed by a solid-state reaction technique, but can only be obtained by the mild oxidation of NaFeAs. \cite{tod10a,goo10a,fri12a} Additionally, deviations from stoichiometry (Na$_{1-y}$Fe$_{2-x}$As$_2$, with $y \approx 0.1$ and $x \approx 0.3$) for the obtained material were suggested.\cite{fri12a} These findings are in striking contrast to stable, stoichiometric KFe$_2$As$_2$, the end member of the Ba$_{1-x}$K$_x$Fe$_2$As$_2$ series.

An important difference between the Ba$_{1-x}$K$_x$Fe$_2$As$_2$ and Ba$_{1-x}$Na$_x$Fe$_2$As$_2$ series is that in addition to the low temperature antiferromagnetic / orthorhombic phase, that is ubiquitous  in Fe-based superconductors, an antiferromagnetic tetragonal, $C4$, phase was reported in the  Ba$_{1-x}$Na$_x$Fe$_2$As$_2$ series over a narrow Na - concentration region around $x \sim 0.25$. \cite{avc13a,avc14a} The tip of the narrow $C4$ dome was suggested to be at $\sim 50$ K; at lower temperatures this new magnetic phase was suggested to co-exist with superconductivity. The bulk physical properties of the members of the series close to and in the $C4$ phase as well as the effect of this magnetic phase on superconductivity are largely unexplored and are interesting to study in detail.

\section{Specific heat capacity and BNC scaling}

\begin{figure} [ht]
\vspace*{8pt}
\centerline{\psfig{file=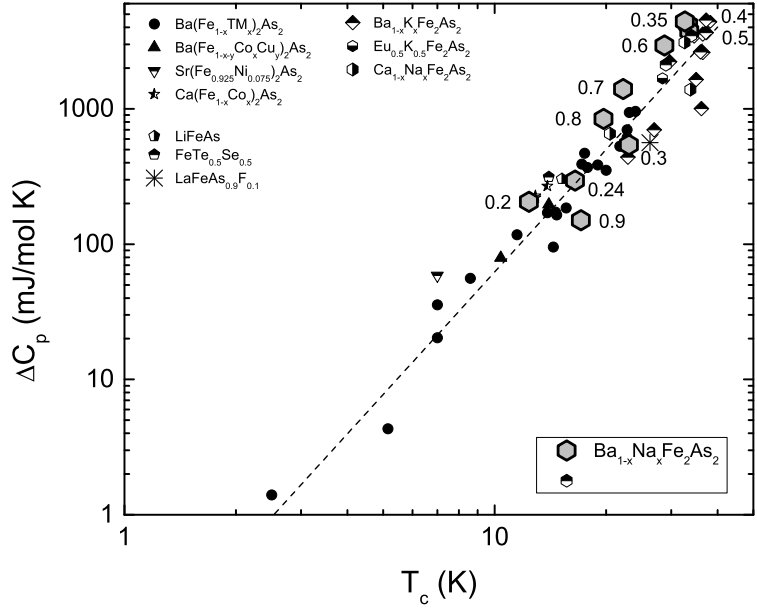,width=10cm}}
\vspace*{8pt}
\caption{$\Delta C_p$ at the superconducting transition vs $T_c$  for the  Ba$_{1-x}$Na$_x$Fe$_2$As$_2$ series, plotted together with literature data for various FeAs-based superconducting materials (after Ref.~\protect\refcite{bud14a}).  Numbers near the symbols are Na - concentrations $x$.  The dashed line corresponds to $\Delta C_p (T_c) \propto T_c^3$. \label{f2}}
\end{figure}

Superconducting samples in both series,  Ba$_{1-x}$K$_x$Fe$_2$As$_2$ ($0.2 \leq x \leq 1$) and  Ba$_{1-x}$Na$_x$Fe$_2$As$_2$ ($0.15 \leq x \leq 0.9$) show a distinct feature in specific heat at $T_c$ \cite{bud13a,bud14a} making determination of $T_c$  and  $\Delta C_p(T_c)$ (by an isoentropic procedure consistent with that used in Ref. \refcite{bud09b}) rather simple. The specific heat capacity jump data for the  Ba$_{1-x}$Na$_x$Fe$_2$As$_2$ series are shown in Fig. \ref{f2} together with the data for a number of Fe - based superconductors from the evolving literature, including data for two closely spaced Na concentrations, $x = 0.35, 0.4$, obtained in  Refs. \refcite{pra11a,asw12a}. There appears to be a clear trend: all data points for Ba$_{1-x}$Na$_x$Fe$_2$As$_2$ follow the BNC scaling. This behavior is similar to previously the studied Ba(Fe$_{1-x}${\it TM}$_x$)$_2$As$_2$ ({\it TM} = transition metal) series, for which the BNC scaling was observed for the samples covering the full extent of the superconducting dome. For the Ba$_{1-x}$Na$_x$Fe$_2$As$_2$ series, $\Delta C_p$ at $T_c$ increases and decreases as $T_c$ rises and falls to form the superconducting dome.

\begin{figure} [ht]
\vspace*{8pt}
\centerline{\psfig{file=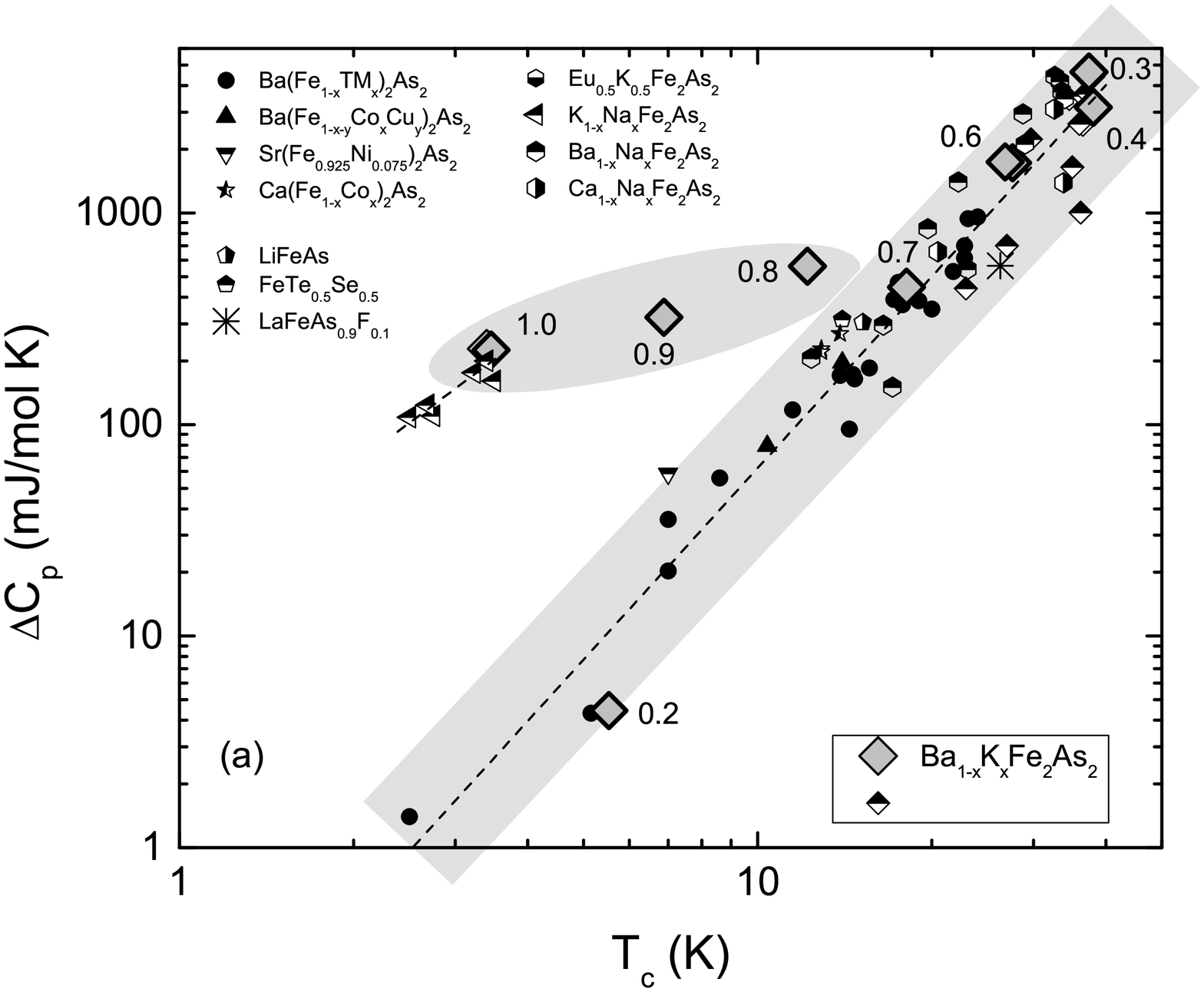,width=9cm}}
\centerline{\psfig{file=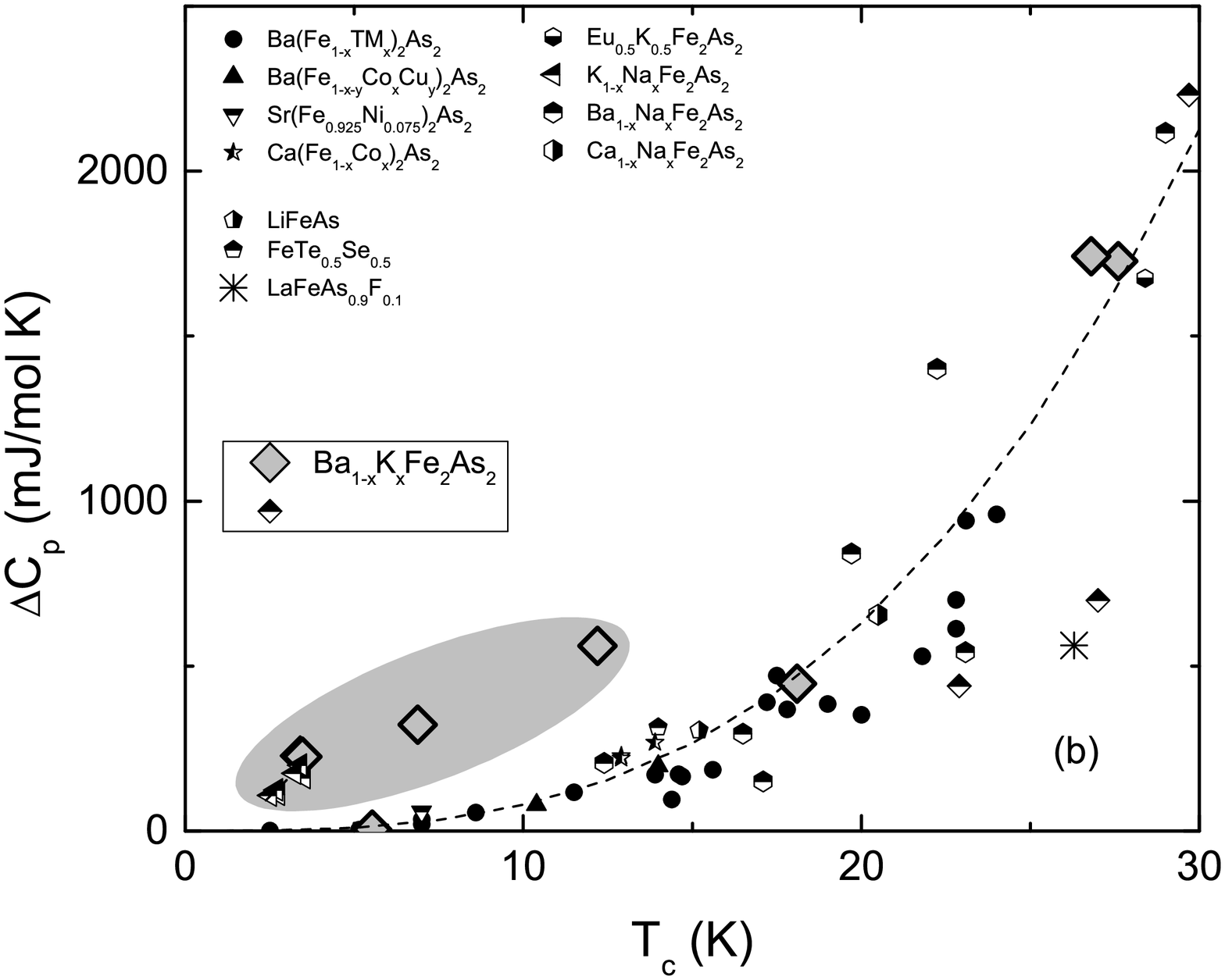,width=9cm}}
\vspace*{8pt}
\caption{(a) $\Delta C_p$ at the superconducting transition vs $T_c$  for the  Ba$_{1-x}$K$_x$Fe$_2$As$_2$ series, plotted together with literature data for various FeAs-based superconducting materials (after Refs.~\protect\refcite{bud13a,bud14a}). Numbers near the symbols are K - concentrations $x$. The dashed line corresponds to $\Delta C_p (T_c)\propto T_c^3$. The short dashed line through K$_{1-x}$Na$_x$Fe$_2$As$_2$ points corresponds to  $\Delta C_p (T_c)\propto T_c^2$. Regions of BNC scaling and deviation of this scaling are highlighted. (b) Low temperature part of the (a) panel plotted on a linear scale. Region of deviation of the BNC scaling is highlighted. \label{f3}}
\end{figure}

The specific heat capacity jump data for  the  Ba$_{1-x}$K$_x$Fe$_2$As$_2$ series obtained in this work were added in Fig. \ref{f3} to the updated BNC plot from Fig. \ref{f2}. There appears to be a clear trend in these data: for $0.2 \leq x \leq 0.7$ the data follow the BNC scaling, in agreement with the scattered literature data for the samples with K - concentrations in this range. The data for $0.8 \leq x \leq 1.0$ clearly deviate from this scaling (as seen both on a {\it log - log} plot and on a linear plot, Fig. \ref{f3}), with the data for the end-compound, KFe$_2$As$_2$, consistent with the other literature values. \cite{bud12a,kim11c} This is clearly different from the Ba(Fe$_{1-x}${\it TM}$_x$)$_2$As$_2$ series, and, even more importantly, from the  Ba$_{1-x}$Na$_x$Fe$_2$As$_2$ series  (closely related to K-doped BaFe$_2$As$_2$) for which the BNC scaling was observed for the samples covering the full extent of the superconducting dome.

Recent data for several  K$_{1-x}$Na$_x$Fe$_2$As$_2$ samples near the K - end of the series ($x \leq 0.31$) \cite{gri14a}, as anticipated  show deviation from the BNC scaling (Fig. \ref{f3}). Within this particular group the $\Delta C_p (T_c)\propto T_c^n$ scaling with $n \approx 2$ was reported. The exponent $n \approx 2$ was explained within a model of two-band {\it d}-wave superconductor with weak pair breaking due to nonmagnetic impurities\cite{gri14a} .

\section{Pressure dependence of the superconducting transition temperature}

\begin{figure} [ht]
\vspace*{8pt}
\centerline{\psfig{file=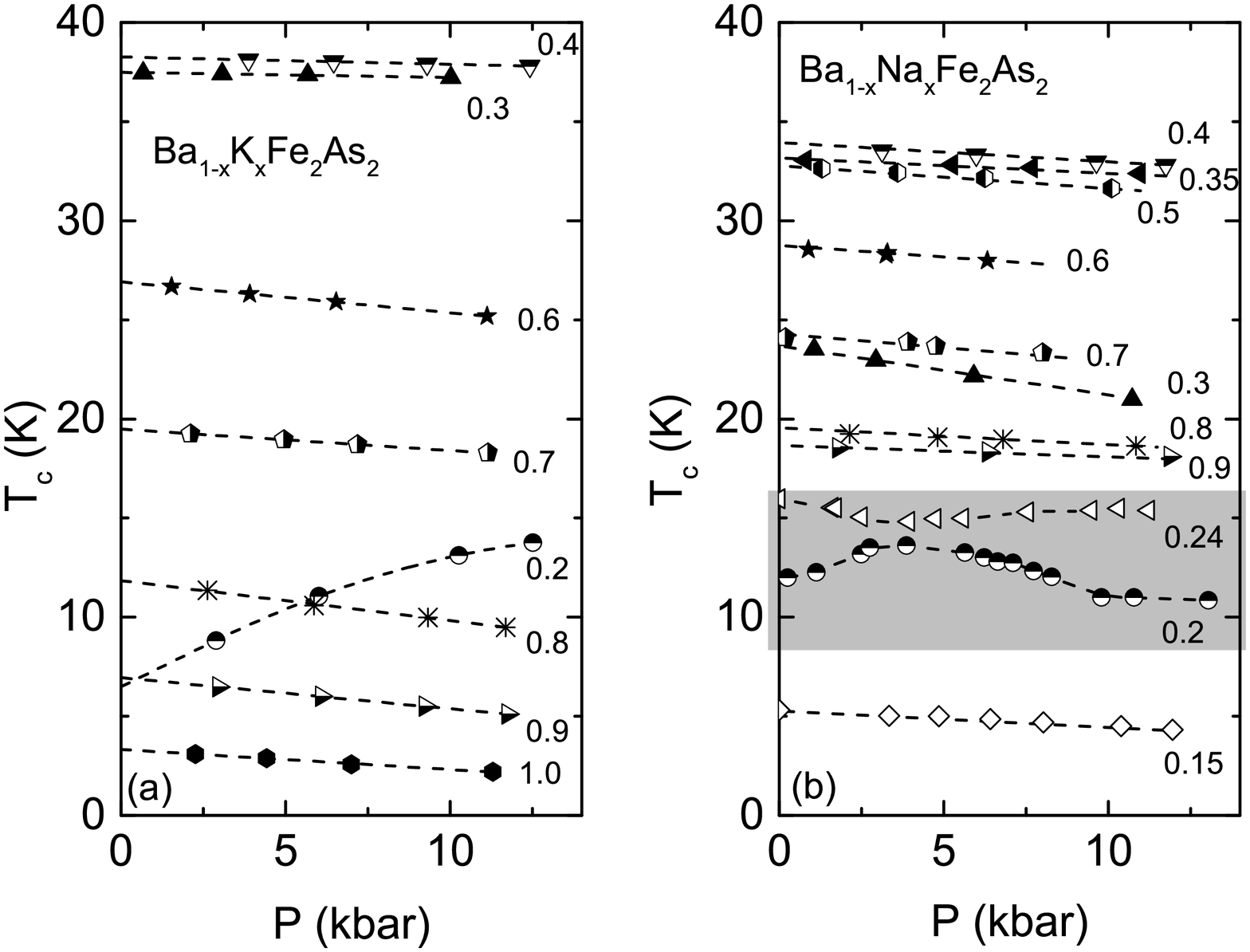,width=12cm}}
\vspace*{8pt}
\caption{Summary plot of the pressure dependence of $T_c$ for the (a) Ba$_{1-x}$K$_x$Fe$_2$As$_2$  and (b) Ba$_{1-x}$Na$_x$Fe$_2$As$_2$ samples (after Refs. ~\protect\refcite{bud13a,bud14a}). Dashed lines are guides to the eye. Data for the Ba$_{1-x}$Na$_x$Fe$_2$As$_2$ samples with $x = 0.20,~0.24$ are highlighted.  \label{f4}}
\end{figure}

The pressure dependencies of the superconducting transition temperatures for the  Ba$_{1-x}$K$_x$Fe$_2$As$_2$ series are shown in Fig. \ref{f4}a \cite{bud13a}. For the underdoped sample, with $x = 0.2$, $T_c$ increases under pressure, for samples close to the optimally doped, $x = 0.3, 0.4$, the $T_c(P)$ dependencies are basically flat, and for the overdoped samples, $x \geq 0.6$, $T_c$ decreases under pressure.  For all samples, except the underdoped $x = 0.2$  the $T_c(P)$ behavior is linear, as such the pressure dependencies of $T_c$ can be well represented by $dT_c/dP$ values. 

\begin{figure} [ht]
\vspace*{8pt}
\centerline{\psfig{file=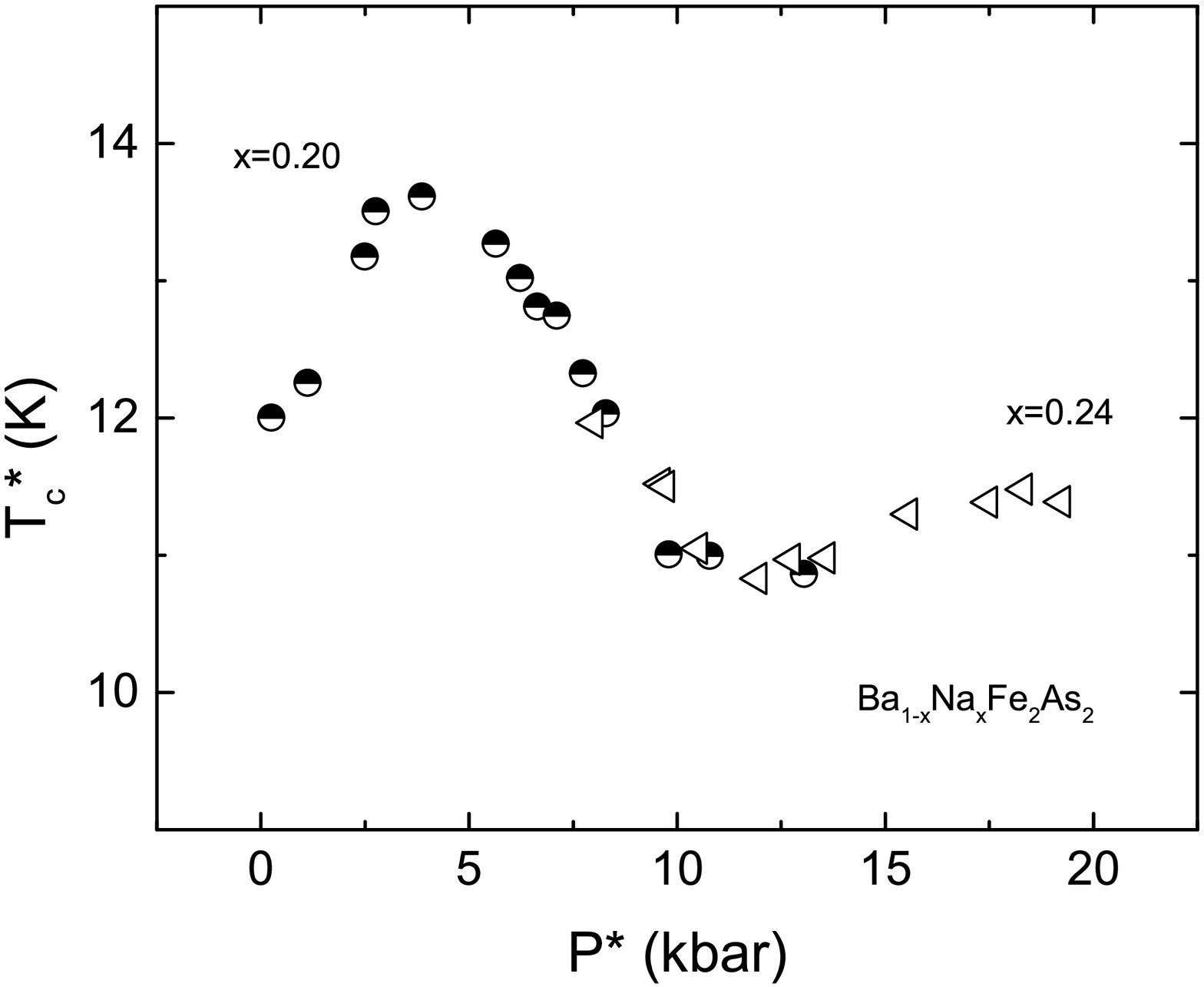,width=10cm}}
\vspace*{8pt}
\caption{Pressure dependence of the superconducting transition temperature for  Ba$_{0.8}$Na$_{0.2}$Fe$_2$As$_2$ and Ba$_{0.76}$Na$_{0.24}$Fe$_2$As$_2$. The data set for $x = 0.24$ is shifted by $+ 8$ kbar along the $X$ - axis and by $- 4$ K along the $Y$ - axis (after Ref.~\protect\refcite{bud14a}). \label{f5}}
\end{figure}

The pressure dependencies of the superconducting transition temperatures of the Ba$_{1-x}$Na$_x$Fe$_2$As$_2$ samples with Na - concentration in the range of $0.15 \leq x \leq 0.9$ are more interesting. They data are shown in Fig. \ref{f4}b \cite{bud14a}. It is noteworthy that (i) $T_c(P)$ are non-monotonic (even in a limited pressure range of this work) for two Na concentrations, $x = 0.2$ and $x = 0.24$; (ii) for all other concentrations studied in this work, both in underdoped and overdoped regimes, $T_c$ decreases under pressure. For  $x = 0.2$ and $x = 0.24$ samples the observed non-monotonic behavior is robust and not affected by pressure cycling. One can join these two data sets  by shifting the data for Ba$_{0.76}$Na$_{0.24}$Fe$_2$As$_2$  by $+ 8$ kbar along the $X$ - axis and by $- 4$ K along the $Y$ - axis (Fig. \ref{f5}).  The grounds for such two-axis shift can be understood if  both, steric effect and hole doping, are assumed to cause changes in superconducting transition temperature.

\begin{figure} [ht]
\vspace*{8pt}
\centerline{\psfig{file=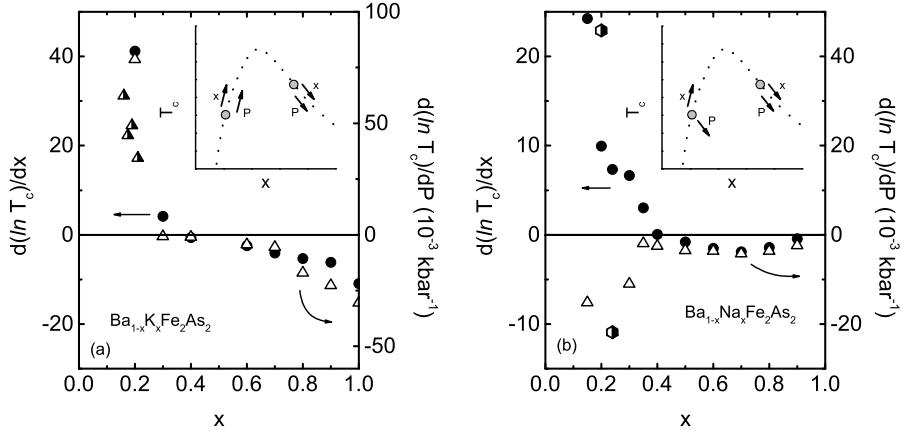,width=12cm}}
\vspace*{8pt}
\caption{ K- [panel (a)] and Na- [panel (b)] concentration dependence of the normalized concentration derivatives, $d(\ln T_c)/dx = \frac{1}{T_{c0}}~d T_c/dx$ (left axis, circles), and the normalized pressure derivatives, $d(\ln T_c)/dP = \frac{1}{T_{c0}}~d T_c/dP$ (right axis, triangles)  of the superconducting transition temperatures for the Ba$_{1-x}$K$_x$Fe$_2$As$_2$ and Ba$_{1-x}$Na$_x$Fe$_2$As$_2$ series respectively (after Refs. ~\protect\refcite{bud13a,bud14a}).  For underdoped Ba$_{1-x}$K$_x$Fe$_2$As$_2$ the initial pressure derivatives extracted from Ref. ~\protect\refcite{has12a} are included (half-filled trangles).  For  Ba$_{0.8}$Na$_{0.2}$Fe$_2$As$_2$ and Ba$_{0.76}$Na$_{0.24}$Fe$_2$As$_2$  the initial, low pressure, normalized pressure derivatives' values are used (half-filled hexagons). Insets: schematic exhibiting {\it similar} effects of substitution and pressure for the Ba$_{1-x}$K$_x$Fe$_2$As$_2$ series [panel (a)]; and {\it different} signs of the change in $T_c$ with substitution and with pressure for the underdoped samples, and {\it similar} effects of substitution and pressure for the overdoped samples for the Ba$_{1-x}$Na$_x$Fe$_2$As$_2$ series [panel (b)]. \label{f6}}
\end{figure}

A more compact way to look at the pressure- and doping- dependence of $T_c$ in the  Ba$_{1-x}$K$_x$Fe$_2$As$_2$ and Ba$_{1-x}$Na$_x$Fe$_2$As$_2$ series is presented in Fig. \ref{f6}, where  the relative changes in superconducting transition temperature under pressure, $d(\ln T_c)/dP$, and with K/Na - doping,  $d(\ln T_c)/dx$ are plotted for comparison. Since the overall shapes of the superconducting domes, $T_c(x)$, in both series are very similar, it is not surprizing that $d(\ln T_c)/dx$ plotted as a function of K or Na concentration (Figs. \ref{f6}a and b, respectively) are comparable as well.

For the  Ba$_{1-x}$K$_x$Fe$_2$As$_2$ series (Fig. \ref{f6}a) the K-concentration dependent pressure derivatives,  $d(\ln T_c)/dP$ show a clear trend and three different K-concentrations regions. The pressure derivatives are positive and rather high for the significantly underdoped samples ($x \sim 0.2$, $T_c < 20$ K). They become small and negative for $0.3 \leq x \leq 0.7$, this concentration range covers slightly underdoped, optimally doped, and part of the overdoped samples.  On further increase of the K-concentration the pressure derivatives continue to be negative with the absolute values, $|d(\ln T_c)/dP|$, increasing as $x$ increases. Fig. \ref{f6}a also allows to compare the relative changes in superconducting transition temperature under pressure and with K - doping.   For $0.2 \leq x \leq 0.7$ both sets of data can be scaled reasonably well, illustrating apparent equivalence of the effect of pressure and doping on $T_c$, suggested for other members of the 122 family. \cite{kim09a,dro10a,kli10a,kim11a} This scaling (in quantative terms) however fails for $0.7 < x \leq 1.0$. Overall,  in the underdoped region both increase in $x$ and  pressure causes increase in $T_c$,  in the optimally doped and overdoped regions both, increase in $x$ and pressure cause decrease in $T_c$ (Fig. \ref{f6}a, inset).

In the Ba$_{1-x}$Na$_x$Fe$_2$As$_2$ series, the initial, low pressure, value of $d(\ln T_c)/dP$ for $x = 0.2$ is positive and relatively high. For other Na - concentrations studied the pressure derivatives of $T_c$ are negative. Whereas for optimally doped and overdoped the absolute values of $d(\ln T_c)/dP$ are rather small and change smoothly with concentration (Fig. \ref{f6}b), there appears to be a break of the trend in the underdoped region. 

A comparison of the relative changes in superconducting transition temperature in the  Ba$_{1-x}$Na$_x$Fe$_2$As$_2$ series  under pressure and with Na - doping is visualized in Fig. \ref{f6}b.   For $0.4 \leq x \leq 0.9$ both sets of data can be scaled reasonably well, illustrating apparent equivalence of the effect of pressure and doping on $T_c$, suggested for other members of the 122 family  and also observed in the limited range of K - concentrations for a closely related  Ba$_{1-x}$K$_x$Fe$_2$As$_2$.  This scaling however fails for Na concentrations $0.15 < x \leq 0.35$. Not just the values of  $d(\ln T_c)/dP$ and  $d(\ln T_c)/dx$ cannot be scaled in this region of concentrations, but (except for $x = 0.2$) the signs of these derivatives are different. In the underdoped region increase in $x$ causes an increase in $T_c$, and pressure causes decrease in $T_c$, however in the optimally doped and overdoped regions both, increase in $x$ and pressure cause decrease in $T_c$ (Fig. \ref{f6}b, inset).

It is noticeable that the absolute values of the normalized pressure derivatives, both  $d(\ln T_c)/dx$ and  $d(\ln T_c)/dP$, for similar values of $x$, appear to be almost factor of 2 higher for  Ba$_{1-x}$K$_x$Fe$_2$As$_2$  than for  Ba$_{1-x}$Na$_x$Fe$_2$As$_2$ .

\section{Discussion}

Both K and Na substitutions in BaFe$_2$As$_2$  provide hole doping and induce superconductivity with comparable maximum values of $T_c$ of $34 - 38$ K at similar K or Na concentrations of $x \approx 0.4$. 

Whereas in the  Ba$_{1-x}$K$_x$Fe$_2$As$_2$ series a clear deviation from the BNC scaling is observed for $0.8 \leq x \leq 1$,  in the  Ba$_{1-x}$Na$_x$Fe$_2$As$_2$ series the data for $0.15 \leq x \leq 0.9$ follow the BNC scaling,  $\Delta C_p(T_c) \propto T_c^3$, fairly well. This probably means that either there is no significant modification of the superconducting state (e.g. change in superconducting gap symmetry) in the Ba$_{1-x}$Na$_x$Fe$_2$As$_2$ series over the whole studied Na concentration range, or, if such modification exists, it is very subtle in its implications for the BNC scaling. The fact that the  Ba$_{1-x}$Na$_x$Fe$_2$As$_2$ series does not extend, in single phase form, to $x = 1.0$ prevents us from carrying this study to pure NaFe$_2$As$_2$, as we were able to do for KFe$_2$As$_2$.

For the  Ba$_{1-x}$K$_x$Fe$_2$As$_2$ series in the $0.7 < x \leq 1$ concentrations interval (the same, where the clear deviations from the BNC scaling of the specific heat capacity jump is observed) the scaling of  $d(\ln T_c)/dP~vs.~x$ and  $d(\ln T_c)/dx~vs.~x$ appears to fail. Although the difference is quantitative rather than qualitative, it might serve as an additional indication of the significant changes in the superconducting state for $x > 0.7$. 

In the recent studies of pure KFe$_2$As$_2$ and CsFe$_2$As$_2$ under pressure \cite{taf13a,taf14a,tau14a,ter14a} a change in the pressure dependence of $T_c$ (combined with an asserted absence of the Lifshitz transition \cite{taf13a,taf14a,ter14a}) was interpreted as a signature of a significant change in the superconducting state: either from $d$-wave to  $s_{\pm}$ \cite{taf13a,taf14a}, or a crossover from a nodal to a full gap $s$-wave superconductivity \cite{ter14a}, or, alternatively, as an indication of appearance of $k_z$ modulation in the superconducting gap \cite{tau14a}, or just of non-monotonic variation of the density of states at the Fermi level without Lifshitz transition and without change of the pairing symmetry \cite{gri14b}.  

Returning back to the  Ba$_{1-x}$K$_x$Fe$_2$As$_2$ series: it appears to be a consensus that in optimally doped samples ($x \approx 0.4$) the symmetry of the superconducting state is  $s_{\pm}$. Then on further K-substitution, application of pressure, or by combination of both perturbations, the system will move towards the direction of pure KFe$_2$As$_2$, so a crossover from $s_{\pm}$ to nodal, or $d$-wave superconductivity is expected to occur at some critical K - concentration or pressure. No sharp feature (similar to the one observed in Refs. \refcite{taf13a,taf14a,tau14a,ter14a}) was observed  for  Ba$_{1-x}$K$_x$Fe$_2$As$_2$ in the $T_c(P)$ or $T_c(x)$ data discussed above. The reason might be related to a rather small pressure range combined with rather large steps in K - concentration in the study discussed above \cite{bud13a}, so that the exact critical value of the pressure/concentration was missed. Additionally, scattering associated with doping could broaden a feature for the intermediate concentrations of K in Ba$_{1-x}$K$_x$Fe$_2$As$_2$, as compared with hydrostatic pressure study in pure, parent compound. It might be an apparent failure of the  $d(\ln T_c)/dP~vs.~x$ and  $d(\ln T_c)/dx$ vs. $x$ scaling, that points to  $0.7 < x < 0.8$ range of concentrations where a change in the nature of the superconducting state occurs. Moreover, whereas in KFe$_2$As$_2$ (and CsFe$_2$As$_2$)  it was argued that there is no significant change of the Fermi surface at the critical pressure, multiple studies, including recent detailed band structure calculations \cite{kha14a} and thermoelectric power measurements \cite{hod14a} point to a Lifshitz transition for $x > 0.7$ in  Ba$_{1-x}$K$_x$Fe$_2$As$_2$ that might complicate the comparison even further. Moreover, just a Lifshitz transition, without any modification of the superconducting state, could, in principle, be responsible for the discussed anomaly in the evolution of the pressure derivatives with concentration, along the same lines as evolution of $dT_c/dP$ was understood in several cuprates \cite{xio92a,cao95a}. 

In the the Ba$_{1-x}$Na$_x$Fe$_2$As$_2$ series, the negative sign of the initial pressure derivatives of $T_c$ for the underdoped samples (except for $x = 0.2$) is a clear indication of the non-equivalence of substitution and pressure in this concentration range, that is different from the gross overall equivalence suggested for other 122 series. \cite{kim09a,dro10a,kli10a,kim11a,bud13a} It has to be noted that for the  Ba(Fe$_{1-x}${\it TM}$_x$)$_2$As$_2$  (at least for {\it TM} concentrations that cover the superconducting dome) \cite{nin09a,rul10a,tha10a} and for Ba$_{1-x}$K$_x$Fe$_2$As$_2$ series \cite{rot08b,avc12a} the concentration dependence of the lattice parameters is monotonic and close to linear.  For the  Ba$_{1-x}$Na$_x$Fe$_2$As$_2$ series, the $a$ lattice parameter decreases with increase of $x$  in almost linear fashion, but the $c$ lattice parameter initially increases and then decreases, with  a maximum  at $x \sim 0.4$ in its dependence of Na - concentration. \cite{avc13a}. Although we do not know which particular structural parameter in the Ba$_{1-x}$Na$_x$Fe$_2$As$_2$ series has the dominant contribution to the pressure dependence of $T_c$, this non-monotonic behavior of $c(x)$ might be responsible for the negative sign of the $dT_c/dP$ for underdoped samples. Detailed structural studies under pressure would be useful for deeper understanding of this problem.

\begin{figure} [ht]
\vspace*{8pt}
\centerline{\psfig{file=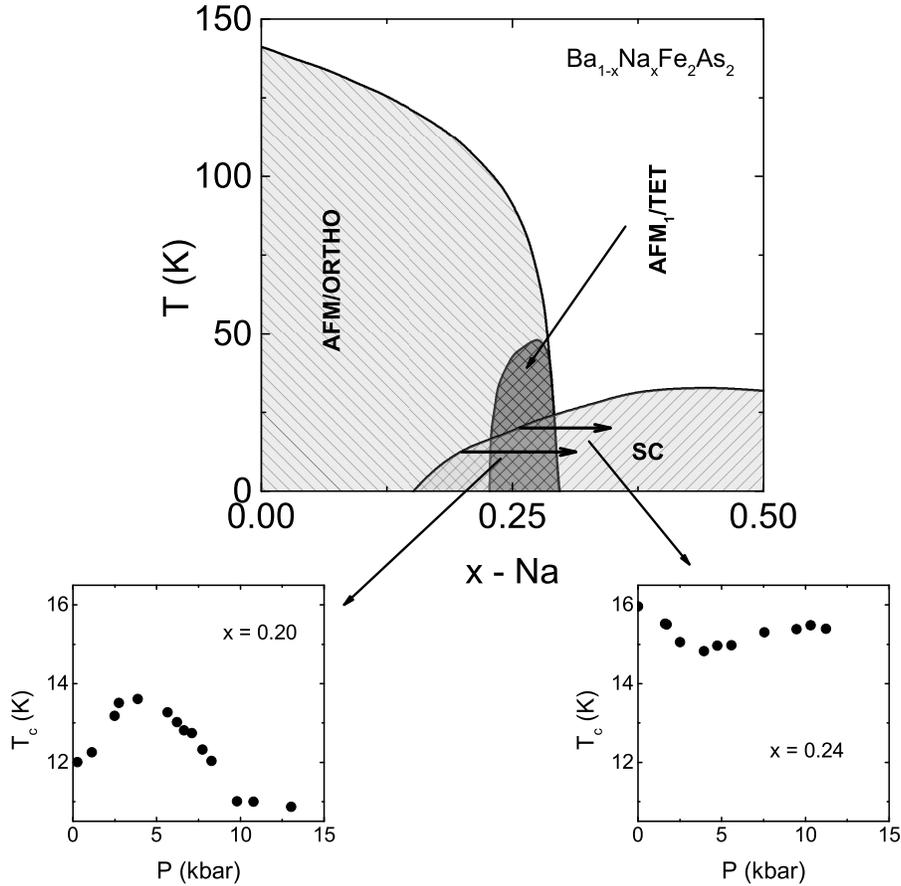,width=12cm}}
\vspace*{8pt}
\caption{Part of the schematic {\it x - T} phase diagram for the  Ba$_{1-x}$Na$_x$Fe$_2$As$_2$ series. The antiferromagnetic/tetragonal, $C4$ phase \protect\cite{avc13a,avc14a} is labeled as {\bf AFM$_1$/TET}.  Horizontal arrows roughly suggest how the phase lines are crossed under pressure for the samples with $x = 0.20$ and $x = 0.24$. Respective $T_c(P)$ plots are reproduced for reference. \label{f7}}
\end{figure}

On one hand, the unusual, non-monotonic behavior of the superconducting transition temperature under pressure for the  Ba$_{1-x}$Na$_x$Fe$_2$As$_2$  samples with $x = 0.2$ and $x = 0.24$ (Fig. \ref{f4}b)  could be considered to be consistent   with what one would expect for a Lifshitz transition \cite{mak65a}. At this moment the data on the Fermi surface of  Ba$_{1-x}$Na$_x$Fe$_2$As$_2$ \cite{asw12a} are regarded as very similar to those for  Ba$_{1-x}$K$_x$Fe$_2$As$_2$ \cite{evt09a,zab08a}, and no change of the Fermi surface topology, from the ARPES measurements, has been reported between the parent, BaFe$_2$As$_2$, compound and  Ba$_{1-x}$Na$_x$Fe$_2$As$_2$ with $x$ values up to 0.4, so the Lifshitz transition hypothesis seems unlikely.

On the other hand, an important feature, unique to the $x - T$ ambient pressure phase diagram of the  Ba$_{1-x}$Na$_x$Fe$_2$As$_2$ series, is the existence of a distinct, narrow antiferromagnetic and tetragonal, $C4$, phase \cite{avc13a,avc14a}.  Both of the samples with non-monotonic pressure dependences of $T_c$ are located, at ambient pressure, close to the phase boundaries of this $C4$ phase. It is thus possible that the observed anomalies in $T_c(P)$ behavior for the  samples with $x = 0.2$ and $x = 0.24$  are associated with the crossing of these phase boundaries under pressure, as shown on the sketch (Fig. \ref{f7}) . If this supposition is correct, the effect of the $C4$ phase on superconductivity requires is different from that of the antiferromagnetic orthorhombic phase that is omnipresent in Fe-As materials, since crossing of the  antiferromagnetic orthorhombic phase line under pressure either has no effect on $T_c$ or this effect has previously been missed. 

Anomalies in the $T_c$ behavior on crossing the $T_N$ line have been observed  in several materials, e.g. with Ir - substitution in Ho(Rh$_{1-x}$Ir$_x$)$_4$B$_4$ \cite{map83a}, or with Dy substitution in Ho$_{1-x}$Dy$_x$Ni$_2$B$_2$C \cite{cho96a}. In CeRhIn$_5$ under pressure the anomaly in $T_c$ was not so clear, but the jump in specific heat capacity at $T_c$ and the initial slope of $H_{c2}$, $dH_{c2}/dT$, had clear features at the pressure corresponding to the crossing of the $T_c$ and $T_N$ phase lines \cite{kne11a}. The exact mechanism and the size of the anomalies might be different, however, at least in the case of Fe - based superconductors, such anomalies ane not unexpected, since antiferromagnetism and superconductivity are competing for the same condiction electrons. Further experimental and theoretical exploration of the details of the superconducting state near, and in co-existence with, an antiferromagnetic order in different systems, including a possibility of effect of magnetism on a symmetry of the superconducting state, would be desirable.

\section{Summary}
The measurements of initial pressure dependencies of $T_c$ and the evolution of the jump in specific heat capacity at $T_c$ (as compared to the BNC scaling) proved to be a useful tool in studies of the iron - based superconductors.  In the  Ba$_{1-x}$K$_x$Fe$_2$As$_2$ series, the results allowed to suggest that significant changes in the nature of the superconducting state concurrent with (or caused by) change of the Fermi surface topology, occur for $x > 0.7$ in the  Ba$_{1-x}$K$_x$Fe$_2$, whereas no significant anomalies were observed in the  Ba$_{1-x}$Na$_x$Fe$_2$As$_2$ series in a similar concentration range.

In the Ba$_{1-x}$Na$_x$Fe$_2$As$_2$ series, it appears that non-monotonic behavior of the $c$ crystallographic lattice parameter and existence of the the narrow antiferromagnetic tetragonal $C4$ phase both affect the pressure dependencies of the superconducting transition temperature. Synthesis of homogeneous single crystals with finely controlled Na concentration  around 20-30\%  and further comprehensive measurements of superconducting and magnetic properties (and their interplay), including studies of the symmetry of the superconducting state,  in this range of concentrations  would be desirable to understand the distinct properties of the  Ba$_{1-x}$Na$_x$Fe$_2$As$_2$ series.

A related,  Ba$_{1-x}$Rb$_x$Fe$_2$As$_2$ family, was reported \cite{pes14a} to have the {\it x - T} phase diagram similar that of  Ba$_{1-x}$K$_x$Fe$_2$As$_2$. The RbFe$_2$As$_2$ end-member is known to be a superconductor, Rb-concentration dependence of the lattice parameters are close to linear, at this moment no $C4$ or any other unusual phase is discovered. This allows us to speculate that the details of the evolution of the Fermi surface and the superconducting state should be very similar to those for the Ba$_{1-x}$Rb$_x$Fe$_2$As$_2$ series. Since the samples covering the whole concentration range exist at this moment only in a polycrystalline form, measurements of $T_c$ under pressure and of the evolution of the jump in specific heat capacity could be a good start, and will probably be done at some point.

Last, but not the least, this this short overview was focused on just two, reasonably simple, types of measurements probing a superconducting state. One should keep in mind, that {\it "La plus belle fille du monde ne peut donner que ce qu'elle a"} and only combination of {\it multiple} experimental techniques and theory has a chance to understand physics of modern complex materials.

\section*{Acknowledgments}

Work at the Ames Laboratory was supported by the US Department of Energy, Basic Energy Sciences, Division of Materials Sciences and Engineering under Contract No. DE-AC02-07CH11358. Fruitful collaboration with Paul C. Canfield, members of the Novel Materials and Ground States research group (Ames Laboratory and Iowa State University) and of the Mercouri G. Kanatzidis research group (Northwestern University and Argonne National Laboratory) is greatly appreciated. A desire to "caress the detail, the divine detail"\footnote{Vladimir V.  Nabokov} contributed to the motivation for this work.


\begin{thebibliography}{0}

\bibitem{joh10a} D. C. Johnston,  {\it Adv. Phys.} {\bf 59} (2010) 803.

\bibitem{ste11a} G. R. Stewart, {\it Rev. Mod. Phys.} {\bf 83} (2011) 1589.

\bibitem{joh11a} Dirk Johrendt, {\it J. Mater. Chem.} {\bf 21} (2011) 13726.

\bibitem{joh11b} Dirk Johrendt, Hideo Hosono, Rolf-Dieter Hoffmann, and Rainer P\"ottgen, {\it Z. Kristallogr.} {\bf 226} (2011) 435.

\bibitem{wan12a} {\it Iron-based Superconductors. Materials, Properties and Mechanisms}, eds.~Nan-Lin Wang, Hideo Hosono, and Pencheng Dai (Pan Stanford Publishing, Boca Raton, FL,  2012)

\bibitem{can10a} Paul C. Canfield and Sergey L. Bud'ko, {\it Annu. Rev. Condens. Matter Phys.} {\bf 1}, 27 (2010).

\bibitem{man10a} D. Mandrus, A. S. Sefat, M. A. McGuire, B. C. Sales, {\it Chem. Mater.} {\bf 22}, 715  (2010).

\bibitem{nin11a} N. Ni and Sergey L. Bud'ko, {\it MRS Bull.} {\bf 36}, 620 (2011).

\bibitem{rot08a} Marianne Rotter, Marcus Tegel, and Dirk Johrendt, {\it Phys. Rev. Lett.} {\bf 101} (2008) 107006.

\bibitem{rot08b} Marianne Rotter, Michael Pangerl, Marcus Tegel, and Dirk Johrendt, {\it Angew. Chem. Int. Ed.} {\bf 47} (2008) 7949.

\bibitem{che09a} H. Chen, Y. Ren, Y. Qiu, Wei Bao, R. H. Liu, G. Wu, T. Wu, Y. L. Xie, X. F. Wang, Q. Huang, and X. H. Chen, {\it EPL} {\bf 85} (2009) 17006.

\bibitem{avc12a} S. Avci, O. Chmaissem, D. Y. Chung, S. Rosenkranz, E. A. Goremychkin, J. P. Castellan, I. S. Todorov, J. A. Schlueter, H. Claus, A. Daoud-Aladine, D. D. Khalyavin, M. G. Kanatzidis, and R. Osborn, {\it Phys. Rev.} {\bf B85} (2012) 184507.

\bibitem{jia09a} Shuai Jiang, Hui Xing, Guofang Xuan, Cao Wang, Zhi Ren, Chumnu Feng, Jianhui Dai, Zhu'an Xu, and Guanghan Cao, {\it J. Phys.: Cond. Mat.} {\bf 21} (2009) 382203.

\bibitem{kas10a} S. Kasahara, T. Shibauchi, K. Hashinoto, K. Ikada, S. Tonegawa, R. Okazaki, H. Shishido, H. Ikeda, H. Takeya, K. Hirata, T. Terashima, and Y. Matsuda, {\it Phys. Rev.} {\bf B81} (2010) 184519.

\bibitem{col09a} E. Colombier, S. L. Bud'ko, N. Ni, and P. C. Canfield, {\it Phys. Rev.} {\bf B79} (2009) 224518.

\bibitem{pag10a} Johnpierre Paglione and Richard L. Greene, {\it Nat. Phys.} {\bf 6} (2010) 645.

\bibitem{bra6569a} N. B. Brandt and N. I. Ginzburg, {\it Usp. Fiz. Nauk} {\bf 85} (1965) 485 [{\it Sov. Phys. Usp.} {\bf 8} (1965) 202]; {\it ibid.} {\bf 98} (1969) 95 [{\it ibid.} {\bf 12} (1969) 344].

\bibitem{lor05a} B. Lorenz and C. W. Chu, in {\it Frontiers in Superconducting Materials}, ed. ~A. V. Narlikar (Springer Berlin Heidelberg, 2005), p. ~459.

\bibitem{sch07a} James S. Schilling,  in: {\it Handbook of High Temperature Superconductivity: Theory and Experiment}, eds. ~J. R. Schrieffer and J. S. Brooks, (Springer Verlag, Hamburg, 2007), p. ~427.

\bibitem{chu09a} C.W. Chu and B. Lorenz, {\it Physica C} {\bf 469} (2009) 385.

\bibitem{sch13a} James Schilling, Narelle Hillier, and Neda Foroozani, {\it J. Phys: Conf. Series} {\bf 449} (2013) 012021.

\bibitem{kim09a} Simon A. J. Kimber, Andreas Kreyssig, Yu-Zhong Zhang, Harald O. Jeschke, Roser Valentí, Fabiano Yokaichiya, Estelle Colombier, Jiaqiang Yan, Thomas C. Hansen, Tapan Chatterji, Robert J. McQueeney, Paul C. Canfield, Alan I. Goldman, and  Dimitri N. Argyriou, {\it Nature Materials} {\bf 8} (2009) 471.

\bibitem{dro10a} Sandra Drotziger, Peter Schweiss, Kai Grube, Thomas Wolf, Peter Adelmann, Christoph Meingast, and Hilbert v. L\"ohneisen, {\it J. Phys. Soc. Jpn.} {\bf 79} (2010) 124705.

\bibitem{kli10a} Lina E. Klintberg, Swee K. Goh,  Shigeru Kasahara, Yusuke Nakai, Kenji Ishida,  Michael Sutherland, Takasada Shibauchi, Yuji Matsuda,  and Takahito Terashima, {\it J. Phys. Soc. Jpn.} {\bf 79} (2010) 123706.

\bibitem{kim11a} S. K. Kim, M. S. Torikachvili, E. Colombier, A. Thaler, S. L. Bud'ko, and P. C. Canfield, Phys. Rev. {\bf B84}, (2011) 134525.

\bibitem{lif60a}  I. M. Lifshitz, {\it Z. Eksp. Teor. Fiz.} {\bf 38} (1960) 1569 [{\it Soviet Phys. JETP} {\bf 11} (1960) 1130]. 

\bibitem{mak65a} V. I. Makarov and V. G. Bar'yakhtar, {\it Zh. Eksp. Teor. Fiz.} {\bf  48} (1965) 1717 [{\it Sov. Phys. JETP} {\bf 21} (1965) 1151].

\bibitem{xio92a} Q. Xiong, J. W. Chu, Y. Y. Sun, H. H. Feng, S. Bud'ko, P. H. Hor, and C. W. Chu, {\it Phys. Rev.} {\bf B46} (1992) 581.

\bibitem{cao95a} Y. Cao, Q. Xiong, Y. Y. Xue, and C. W. Chu, {\it Phys. Rev.} {\bf B52} (1995) 6854.

\bibitem{fie96a} W. H. Fietz, R. Quenzel, H. A. Ludwig, K. Grube, S. I. Schlachter, F. W. Hornung, T. Wolf, A. Erb, M. Kl\"aser, G. M\"uller-Vogt, {\it Physica C} {\bf 270} (1996) 258.

\bibitem{taf13a} F. F. Tafti, A. Juneau-Fecteau, M.-\`E. Delage, S. Ren\'e de Cotret, J.-Ph. Reid, A. F. Wang, X.-G. Luo, X. H. Chen, N. Doiron-Leyraud, and Louis Taillefer, {\it Nat. Phys.}  {\bf 9} (2013) 349.

\bibitem{taf14a} F. F. Tafti, J. P. Clancy, M. Lapointe-Major, C. Collignon, S. Faucher, J. A. Sears, A. Juneau-Fecteau, N. Doiron-Leyraud, A. F. Wang, X.-G. Luo, X. H. Chen, S. Desgreniers, Young-June Kim, and Louis Taillefer, {\it Phys. Rev.} {\bf B89} (2014) 134502.

\bibitem{tau14a} Valentin Taufour, Neda Foroozani, Makariy A. Tanatar, Jinhyuk Lim, Udhara Kaluarachchi, Stella K. Kim, Yong Liu, Thomas A. Lograsso, Vladimir G. Kogan, Ruslan Prozorov, Sergey L. Bud'ko, James S. Schilling, and Paul C. Canfield, {\it Phys. Rev.} {\bf B89} (2014) 220509.

\bibitem{car90a} J. P. Carbotte, {\it Rev. Mod. Phys.} {\bf 62} (1990) 1027.

\bibitem{bou01a} F. Bouquet, Y. Wang, R. A. Fisher, D. G. Hinks, J. D. Jorgensen, A. Junod, and N. E. Phillips, {\it EPL} {\bf 56} (2001) 856.

\bibitem{joh13a} David C. Johnston, {\it Supercond. Sci. Technol.} {\bf 26} (2013) 115011.

\bibitem{ska64a}  S. Skalski, O. Betbeder-Matibet, and P. R. Weiss, {\it Phys. Rev.} {\bf 136} (1964) A1500.

\bibitem{mul72a} E. M\"uller-Hartmann and J. Zittartz, {\it Solid State Commun.} {\bf 11} (1972) 401.

\bibitem{map76a} M. Brian Maple, {\it Appl. Phys.} {\bf 9} (1976) 179. 

\bibitem{shi73a} H. Shiba, {\it Prog. Theor. Phys.}  {\bf 50} (1973) 50.

\bibitem{ope04a} L. A. Openov, {\it Phys. Rev.}  {\bf B69} (2004) 224516.

\bibitem{gur11a} Alex Gurevich, {\it Nat. Mat.} {\bf 10} (2011) 255.

\bibitem{bud09b} S. L. Bud'ko, N. Ni, and P. C. Canfield,  {\it Phys. Rev.} {\bf B79} (2009) 220516.

\bibitem{kim11b} J. S. Kim, G. R. Stewart, S. Kasahara, T. Shibauchi, T. Terashima, and Y. Matsuda, {\it J. Phys.: Condens. Matter} {\bf 23} (2011) 222201.

\bibitem{kim12a} J. S. Kim, B. D. Faeth, and G. R. Stewart, {\it Phys. Rev.} {\bf B86} (2012) 054509.

\bibitem{zaa09a} J. Zaanen, {\it Phys. Rev.} {\bf B80} (2009) 212502. 

\bibitem{kog09a} V. G. Kogan, {\it Phys. Rev.} {\bf B80} (2009) 214532; {\it ibid.} {\bf B81} (2010) 184528. 

\bibitem{vav11a} M. G. Vavilov, A. V. Chubukov, and A. B. Vorontsov, {\it Phys. Rev.} {\bf B84} (2011) 140502 (2011); M. G. Vavilov and A. V. Chubukov, {\it ibid.} {\bf B84}, (2011) 214521.

\bibitem{cor10a} Raquel Cortes-Gil, Dinah R. Parker, Michael J. Pitcher, Joke Hadermann, and Simon J. Clarke, {\it Chem Mater.} {\bf 22} (2010) 4304.

\bibitem{avc13a} S. Avci, J. M. Allred, O. Chmaissem, D. Y. Chung, S. Rosenkranz, J. A. Schlueter, H. Claus, A. Daoud-Aladine, D. D. Khalyavin, P. Manuel, A. Llobet, M. R. Suchomel, M. G. Kanatzidis, and R. Osborn, {\it Phys. Rev.} {\bf B88} (2013) 094510.

\bibitem{avc14a} S. Avci, O. Chmaissem, J. M. Allred, S. Rosenkranz, I. Eremin, A. V. Chubukov, D. E. Bugaris, D. Y. Chung, M. G. Kanatzidis, J.- P Castellan, J. A. Schlueter, H. Claus 	D. D. Khalyavin, P. Manuel, A. Daoud-Aladine, and R. Osborn, {\it Nat. Comm.} {\bf 5} (2014) 3845.

\bibitem{nin08a} N. Ni, M. E. Tillman, J.-Q. Yan, A. Kracher, S. T. Hannahs, S. L. Bud'ko, and P. C. Canfield, {\it Phys. Rev.} {\bf B78} (2008) 214515.

\bibitem{wie11a} Erwin Wiesenmayer, Hubertus Luetkens, Gwendolyne Pascua, Rustem Khasanov, Alex Amato, Heidi Potts, Benjamin Banusch, Hans-Henning Klauss, and Dirk Johrendt, {\it Phys. Rev. Lett.} {\bf 107}  (2011) 237001.

\bibitem{pra09a} D. K. Pratt, W. Tian, A. Kreyssig, J. L. Zarestky, S. Nandi, N. Ni, S. L. Bud'ko, P. C. Canfield, A. I. Goldman, and R. J. McQueeney, {\it Phys. Rev. Lett.} {\bf 103} (2009) 087001.

\bibitem{din09a} H. Ding, P. Richard, K. Nakayama, K. Sugawara, T. Arakane, Y. Sekiba, A. Takayama, S. Souma, T. Sato, T. Takahashi, Z. Wang, X. Dai, Z. Fang, G. F. Chen, J. L. Luo, and N. L. Wang, {\it EPL} {\bf 83} (2008) 47001.

\bibitem{chr08a} A. D. Christianson, E. A. Goremychkin, R. Osborn, S. Rosenkranz, M. D. Lumsden, C. D. Malliakas, I. S. Todorov, H. Claus, D. Y. Chung, M. G. Kanatzidis, R. I. Bewley, and T. Guidi, {\it Nature} {\bf 456} (2008) 930.

\bibitem{luo09a} X. G. Luo, M. A. Tanatar,  J.-Ph. Reid, H. Shakeripour, N. Doiron-Leyraud, N. Ni, S. L. Bud'ko, P. C. Canﬁeld, Huiqian Luo, Zhaosheng Wang, Hai-Hu Wen, R. Prozorov, and Louis Taillefer, {\it Phys. Rev.} {\bf B80} (2009) 140503.

\bibitem{gra09a} S. Graser, T. A. Maier, P. J. Hirschfeld, and D. J. Scalapino, {\it New J. Phys.} {\bf 11} (2009) 025016.

\bibitem{maz09a} I. I. Mazin and J. Schmalian, {\it Physica C} {\bf 469} (2009) 614.

\bibitem{nak11a} K. Nakayama, T. Sato, P. Richard, Y.-M. Xu, T. Kawahara, K. Umezawa, T. Qian, M. Neupane, G. F. Chen, H. Ding, and T. Takahashi, {\it Phys. Rev.} {\bf B83} (2011) 020501.

\bibitem{ota14a} Y. Ota, K. Okazaki, Y. Kotani, T. Shimojima, W. Malaeb, S. Watanabe, C.-T. Chen, K. Kihou, C. H. Lee, A. Iyo, H. Eisaki, T. Saito, H. Fukazawa, Y. Kohori, and S. Shin,  {\it Phys. Rev.} {\bf B89} (2014) 081103.

\bibitem{sat09a} T. Sato, K. Nakayama, Y. Sekiba, P. Richard, Y. - M. Xu, S. Souma, T. Takahashi, G. F. Chen, J. L. Luo, N. L. Wang, and H. Ding, {\it Phys. Rev. Lett.} {\bf 103} (2009) 047002.

\bibitem{has10a} K. Hashimoto, A. Serafin, S. Tonegawa, R. Katsumata, R. Okazaki, T. Saito, H. Fukazawa, Y. Kohori, K. Kihou, C. H. Lee, A. Iyo, H. Eisaki, H. Ikeda, Y. Matsuda, A. Carrington, and T. Shibauchi, {\it Phys. Rev.} {\bf B82} (2010) 014526.

\bibitem{don10a} J. K. Dong, S. Y. Zhou, T. Y. Guan, H. Zhang, Y. F. Dai, X. Qiu, X. F. Wang, Y. He, X. H. Chen, and S. Y. Li, {\it Phys. Rev. Lett.} {\bf 104} (2010) 087005.

\bibitem{ter10a} Taichi Terashima, Motoi Kimata, Nobuyuki Kurita, Hidetaka Satsukawa, Atsushi Harada, Kaori Hazama, Motoharu Imai, Akira Sato, Kunihiro Kihou, Chul-Ho Lee, Hijiri Kito, Hiroshi Eisaki, Akira Iyo, Taku Saito, Hideto Fukazawa, Yoh Kohori, Hisatomo Harima, and Shinya Uji, {\it Phys. Rev. Lett.} {\bf 104} (2010) 259701.

\bibitem{don10b} J. K. Dong and S. Y. Li, {\it Phys. Rev. Lett.} {\bf 104} (2010) 259702.

\bibitem{rei12a} J. - Ph. Reid, M. A. Tanatar, A. Juneau-Fecteau, R. T. Gordon, S. Ren\'e de Cotret, N. Doiron-Leyraud, T. Saito, H. Fukazawa, Y. Kohori, K. Kihou, C. H. Lee, A. Iyo, H. Eisaki, R. Prozorov, and Louis Taillefer, {\it Phys. Rev. Lett.} {\bf 109} (2012) 087001.

\bibitem{mai12a} S. Maiti, M. M. Korshunov, and A. V. Chubukov, {\it Phys. Rev.} {\bf B85} (2012) 014511.

\bibitem{tho11a} R. Thomale, C. Platt, W. Hanke, J. Hu, and B. A. Bernevig, {\it Phys. Rev. Lett.} {\bf 107} (2011) 117001.

\bibitem{oka12a} K. Okazaki, Y. Ota, Y. Kotani, W. Malaeb, Y. Ishida, T. Shimojima, T. Kiss, S. Watanabe, C.-T. Chen, K. Kihou, C. H. Lee, A. Iyo, H. Eisaki, T. Saito, H. Fukazawa, Y. Kohori, K. Hashimoto, T. Shibauchi, Y. Matsuda, H. Ikeda, H. Miyahara, R. Arita, A. Chainani, and S. Shin, {\it Science} {\bf 337} (2012) 1314.

\bibitem{rei12b} J.-Ph. Reid, A. Juneau-Fecteau, R. T. Gordon, S. Rene de Cotret, N. Doiron-Leyraud, X. G. Luo, H. Shakeripour, J. Chang, M. A. Tanatar, H. Kim, R. Prozorov, T. Saito, H. Fukazawa, Y. Kohori, K. Kihou, C. H. Lee, A. Iyo, H. Eisaki, B. Shen, H.-H. Wen, and Louis Taillefer, {\it Supercond. Sci. Technol.} {\bf  25} (2012) 084013.

\bibitem{hir12a} Masanori Hirano, Yuji Yamada, Taku Saito, Ryo Nagashima, Takehisa Konishi, Tatsuya Toriyama, Yukinori Ohta,  Hideto Fukazawa, Yoh Kohori, Yuji Furukawa, Kunihiro Kihou, Chul-Ho Lee, Akira Iyo, and Hiroshi Eisaki, {\it J. Phys. Soc. Jpn.} {\bf 81} (2012) 054704.

\bibitem{shi08a} Parasharam Maruti Shirage, Kiichi Miyazawa, Hijiri Kito, Hiroshi Eisaki, and Akira Iyo, {\it Appl. Phys. Express} {\bf 1} (2008) 081702.

\bibitem{wug08a} G. Wu, H. Chen, T. Wu, Y. L. Xie, Y. J. Yan, R. H. Liu, X. F. Wang, J. J. Ying, and X. H. Chen, {\it J. Phys.: Cond. Mat.} {\bf 20} (2008) 422201.

\bibitem{zha11a} K. Zhao, Q. Q. Li, X. C. Wang, Z. Deng, Y. X. Lv, J. L. Zhu, F. Y. Li, and C. Q. Jin, {\it Phys. Rev.} {\bf B84} (2011) 184534.

\bibitem{gok09a} T. Goko, A. A. Aczel, E. Baggio-Saitovitch, S. L. Bud'ko, P. C. Canfield, J. P. Carlo, G. F. Chen, Pengcheng Dai, A. C. Hamann, W. Z. Hu, H. Kageyama, G. M. Luke, J. L. Luo, B. Nachumi, N. Ni, D. Reznik, D. R. Sanchez-Candela, A. T. Savici, K. J. Sikes, N. L. Wang, C. R. Wiebe, T. J. Williams, T. Yamamoto, W. Yu, and Y. J. Uemura, {\it Phys. Rev.} {\bf B80} (2009) 024508. 

\bibitem{cor11a} Raquel Cortes-Gil and Simon J. Clarke, {\it Chem. Mater.} {\bf 23} (2011) 1009.

\bibitem{tod10a} Iliya Todorov, Duck Young Chung, Helmut Claus, Christos D. Malliakas, Alexios P. Douvalis, Thomas Bakas, Jiaqing He, Vinayak P. Dravid, and Mercouri G. Kanatzidis, {\it Chem. Mater.} {\bf 22} (2010) 3916.

\bibitem{goo10a} M. Gooch, B. Lv, K. Sasmal, J. H. Tapp, Z. J. Tang, A. M. Guloy, B. Lorenz, and C. W. Chu, {\it Physica C} {\bf 70} (2010) S276.

\bibitem{fri12a} Gina M. Friederichs, Inga Schellenberg, Rainer P\"ottgen, Viola Duppel, Lorenz Kienle, J\"orn Schmedt auf der G\"unne, and Dirk Johrendt, {\it Inorg. Chem.} {\bf 51} (2012) 8161.

\bibitem{bud13a} Sergey L. Bud'ko, Mihai Sturza, Duck Young Chung, Mercouri G. Kanatzidis, and Paul C. Canfield, {\it Phys. Rev.} {\bf  B87} (2013) 100509.

\bibitem{bud14a} Sergey L. Bud'ko, Duck Young Chung, Daniel Bugaris, Helmut Claus, Mercouri G. Kanatzidis, and Paul C. Canfield, {\it Phys. Rev.} {\bf B89} (2014) 014510.

\bibitem{pra11a} A. K. Pramanik, M. Abdel-Hafiez, S. Aswartham, A. U. B. Wolter, S. Wurmehl, V. Kataev, and B. B\"uchner, {\it Phys. Rev.} {\bf B84} (2011) 064525.

\bibitem{asw12a} S. Aswartham, M. Abdel-Hafiez, D. Bombor, M. Kumar, A. U. B. Wolter, C. Hess, D. V. Evtushinsky, V. B. Zabolotnyy, A. A. Kordyuk, T. K. Kim, S. V. Borisenko, G. Behr, B. B\"uchner, and S. Wurmehl, {\it Phys. Rev.} {\bf B85} (2012) 224520.

\bibitem{bud12a} Sergey L. Bud'ko, Yong Liu, Thomas A. Lograsso, and Paul C. Canfield, {\it Phys. Rev.} {\bf B86} (2012) 224514.

\bibitem{kim11c} J. S. Kim, E. G. Kim, G. R. Stewart, X. H. Chen, and X. F. Wang, {\it Phys. Rev.} {\bf B83} (2011) 172502.

\bibitem{gri14a} V. Grinenko, D. V. Efremov, S.-L. Drechsler, S. Aswartham, D. Gruner, M. Roslova, I. Morozov, K. Nenkov, S. Wurmehl, A. U. B. Wolter, B. Holzapfel, and B. B\"uchner, {\it Phys. Rev.} {\bf B89} (2014) 060504.

\bibitem{has12a} E. Hassinger, G. Gredat, F. Valade, S. Ren\'e de Cotret, A. Juneau-Fecteau, J.-Ph. Reid, H. Kim, M. A. Tanatar, R. Prozorov, B. Shen, H.-H. Wen, N. Doiron-Leyraud, and Louis Taillefer, {\it Phys. Rev.} {\bf B86} (2012) 140502.

\bibitem{ter14a} Taichi Terashima, Kunihiro Kihou, Kaori Sugii, Naoki Kikugawa, Takehiko Matsumoto, Shigeyuki Ishida, Chul-Ho Lee, Akira Iyo, Hiroshi Eisaki, and Shinya Uji, {\it Phys. Rev.} {\bf  B89} (2014) 134520.

\bibitem{gri14b} V. Grinenko, W. Schottenhamel, A. U. B. Wolter, D. V. Efremov, S.-L. Drechsler, S. Aswartham, M. Kumar, S. Wurmehl, M. Roslova, I. V. Morozov, B. Holzapfel, B. B\"uchner, E. Ahrens, S. I. Troyanov, S. Köhler, E. Gati, S. Knöner, N. H. Hoang, M. Lang, F. Ricci, and G. Profeta, {\it Phys. Rev.} {\bf  B90} (2014) 094511.

\bibitem{kha14a} Suffian N. Khan and Duane D. Johnson, {\it Phys. Rev. Lett.} {\bf 112} (2014) 156401.

\bibitem{hod14a} Halyna Hodovanets, Yong Liu, Anton Jesche, Sheng Ran, Eun Deok Mun, Thomas A. Lograsso, Sergey L. Bud'ko, and Paul C. Canfield, {\it Phys. Rev.} {\bf B89} (2014) 224517.

\bibitem{nin09a} N. Ni, A. Thaler, A. Kracher, J. Q. Yan, S. L. Bud'ko, and P. C. Canﬁeld, {\it Phys. Rev.} {\bf B80} (2009) 024511.

\bibitem{rul10a} F. Rullier-Albenque, D. Colson, A. Forget, P. Thu\'ery, and S. Poissonnet, {\it Phys. Rev.} {\bf B81} (2010) 224503.

\bibitem{tha10a} A. Thaler, N. Ni, A. Kracher, J. Q. Yan, S. L. Bud'ko, and P. C. Canﬁeld, {\it Phys. Rev.} {\bf B82} (2010) 014534.

\bibitem{evt09a} D. V. Evtushinsky, D. S. Inosov, V. B. Zabolotnyy, A. Koitzsch, M. Knupfer, B. B\"uchner, M. S. Viazovska, G. L. Sun, V. Hinkov, A. V. Boris,  C. T. Lin, B. Keimer, A. Varykhalov, A. A. Kordyuk, and S. V. Borisenko,  {\it Phys. Rev.} {\bf B79} (2009) 054517.

\bibitem{zab08a} V. B. Zabolotnyy, D. S. Inosov, D. V. Evtushinsky, A. Koitzsch, A. A. Kordyuk, G. L. Sun, J. T. Park, D. Haug, V. Hinkov, A. V. Boris, C. T. Lin, M. Knupfer, A. N. Yaresko, B. B\"uchner, A. Varykhalov, R. Follath, and S. V. Borisenko, {\it Nature (London)} {\bf 457} (2008) 569.

\bibitem{map83a} M. B. Maple, {\it J. Magn. Magn. Mat.} {\bf  31-34} (1983) 479.

\bibitem{cho96a} B. K. Cho, P. C. Canfield, and D. C. Johnston, {\it Phys. Rev. Lett.} {\bf 77} (1996) 163.

\bibitem{kne11a} Georg Knebel, Jonathan Buhot, Dai Aoki, Gerard Lapertot, Stephane Raymond, Eric Ressouche, and Jacques Flouquet, {\it J. Phys. Soc. Jpn.} {\bf 80} (2011) SA001.

\bibitem{pes14a} Simon Peschke, Tobias St\"urzer, and Dirk Johrendt, {\it Z. Anorg. Allg. Chem.} {\bf 640} (2014) 830.


\end{thebibliography}
\end{document}